\documentclass[twocolumn,showpacs,preprintnumbers,superscriptaddress,amsmath,amssymb,floatfix,aps]{revtex4}

\usepackage[dvips]{graphicx}% Include figure files
\usepackage{dcolumn}% Align table columns on decimal point
\usepackage{bm}% bold math
\begin{document}

\title{Application of the Lifshitz theory to poor conductors}

\author{Vitaly B. Svetovoy}
\affiliation{MESA+ Research Institute, University of Twente, PO 217,
7500 AE Enschede, The Netherlands}

\date{\today}

\begin{abstract}
The Lifshitz formula for the dispersive forces is generalized to
the materials, which cannot be described with the local dielectric
response. Principal nonlocality of poor conductors is related with
the finite screening length of the penetrating field and the
collisional relaxation; at low temperatures the role of collisions
plays the Landau damping. The spatial dispersion makes the theory
self consistent. Our predictions are compared with the recent
experiment. It is demonstrated that at low temperatures the
Casimir-Lifshitz entropy disappears as $T$ in the case of
degenerate plasma and as $T^2$ for the nondegenerate one.
\end{abstract}

\maketitle

The Casimir-Lifshitz force is a dispersion interaction of
electromagnetic origin acting between neutral bodies without
permanent polarizations. The original Casimir formula \cite{Cas48}
for the force between ideal metals was extended to real materials by
Lifshitz and coworkers \cite{Lif55,DLP,LP9}. Recently there was
considerable progress in the experimental verification of the theory
(see review \cite{Cap07} and references therein) and various
applications to nanosciences were discussed \cite{Cap07}. The
Lifshitz theory is extensively accepted as a common tool to deal
with dispersive forces in physics, biology, chemistry, and
technology.

The zero temperature contribution to the force originating from
quantum fluctuations of the electromagnetic field is well
understood. On the contrary, the classical or high temperature part
of the thermal contribution (no dependence on $\hbar$) is the source
of constant controversies. There is no continuous transition for the
forces between ideal metals and real metals \cite{Bos00}. It is
related with the transparency of real metals for s-polarized
(transverse electric) low frequency field. It was found that for
real metals there is a thermodynamic problem \cite{Bez02}: the
Casimir-Lifshitz entropy is not going to zero at $T\rightarrow 0$.
This problem is still in debate (for the latest publications see
\cite{Nernst}) but a new controversy for poor conductors emerged
\cite{Gey05}. In this case the reflection coefficient for
p-polarization (transverse magnetic) is discontinuous in the
transition from zero to arbitrarily small conductivity. This
discontinuity again breaks the Nernst heat theorem. Obvious
contradiction to the common sense shows that some important physics
is missed.

In this paper we demonstrate that account for the spatial dispersion
of the materials resolves the problems and make the theory self
consistent. We formulate the condition at which the Lifshitz formula
can be extended to the description of the forces between nonlocal
materials. Special attention is paid to the case of poor conductors
because the nonlocal effects are more important for them than for
dielectrics or good metals.

For dielectrics the nonlocality can be neglected due to the absence
of free charges except maybe the range near polariton resonances.
For metals the local approximation is good because of the very short
Thomas-Fermi screening length. Spatial dispersion for metals is
important at low temperatures when the mean free path for electrons
becomes larger than the field penetration depth and the anomalous
skin effect plays role \cite{Sve03,Ser05,Sve05}.

Recently it was demonstrated by Pitaevskii \cite{Pit08} that the
classical part of the Casimir-Polder force between a bad conductor
and an atom is essentially nonlocal due to the finite screening
length of free charges (Debye length). In Ref. \cite{Pit08} the
force was found in the large distance limit. The formula for the
force interpolates between good metals and dielectrics. It depends
on the density of free carriers via the Debye length $l_D$.

The most general theory allowing the spatial dispersion was
developed by Barash and Ginzburg \cite{Bar75}. The force was found
to have an additional term in comparison with the Lifshitz formula
related with the nonlocal material response. On this bases it was
concluded in Ref. \cite{Kli07} that the Lifshitz formula cannot be
used in the nonlocal case. However, this conclusion does not follow
from \cite{Bar75}. The Lifshitz formula can be applied at least for
plasma-like media. It was demonstrated \cite{Bar75} that if the long
range interaction is taken into account in the plasma dielectric
functions to the first order in the coupling constant $e^2$, then
the Lifshitz formula holds true.

Below it is shown that the dielectric functions of the plasmas in
metals and semiconductors calculated in the random phase
approximation (RPA) can be used with the Lifshitz formula and give
sufficient description of these materials for evaluation of the
Casimir-Lifshitz force.

To account for the nonlocal response of metals it is possible
\cite{Esq04,Ser05} to use the RPA (electrons are independent but
respond not to the external field but to the screened one
\cite{Ash76}). This simple choice is good in the weak-coupling
regime (rarefied plasma), but the plasmas in real metals are
actually strongly coupled. However, significant deviations from RPA
appear at large wavenumbers $k\sim k_F$ \cite{Gor80}, where $k_F$ is
the Fermi wavenumber. For the bodies separated by the distance $a$
the important wavenumbers are $k\sim 1/a\ll k_F$. For this reason
the use of the RPA dielectric functions is justified.

We can use the same approach to construct the nonlocal response of
nondegenerate semiconductors. The difference with metals is that now
electrons (holes) have to be considered as nondegenerate plasma and
instead of the Fermi-Dirac distribution the Boltzmann distribution
can be used. Technically the problem is equivalent to the
weak-coupled nondegenerate plasma, for which one can use the
textbook result \cite{LP10}.

Generalizing the situation we can consider two classes of plasmas:
degenerate and nondegenerate. The first class (I) describes
materials for which density of free charged particles stays finite
at $T=0$. Metals, semimetals, degenerate semiconductors etc. belong
to this class. Materials with the energy gap, for which density of
free charges disappears with $T$ belong to the second class (II).
Representatives of this class are nondegenerate semiconductors,
ionic conductors, many disordered materials etc.

In the nonlocal case the dielectric function becomes a tensor, which
has two independent components: transversal $\varepsilon_t$ and
longitudinal $\varepsilon_l$ with respect to the wave vector ${\bf
k}$. For very small relaxation frequency $\gamma\rightarrow 0$
(collisionless plasma) in RPA these components are
\begin{equation}\label{etl}
    \varepsilon_{t,l}(\omega,k)=\varepsilon_0(\omega)-\frac{\omega_p^2}{\omega(\omega+i\gamma)}
    f_{t,l}(x).
\end{equation}
Here the first term $\varepsilon_0(\omega)$ is introduced to account
for the interband transitions since these processes are beyond the
(quasi)free electron model. The variable $x$ is
\begin{equation}\label{xdef}
    x=\frac{\omega+i\gamma}{v k},\ \ \
    v=\left\{
    \begin{array}{c}
    v_F,\ {\rm for}\ I,\\
    \sqrt{\frac{2T}{m^*}},\ {\rm for}\ II,
    \end{array}
    \right.
\end{equation}
where $m^*$ is the effective mass of the charge carrier (electron or
hole) and $v_F$ is the Fermi velocity. The plasma frequency
$\omega_p$ in (\ref{etl}) and the Debye wavenumber $k_D$ (see below)
are defined as
\begin{equation}\label{omp}
    \omega_p=\sqrt{\frac{4\pi e^2 N}{m^*}}, \ \ \ k_D=\left\{
    \begin{array}{c}
    \sqrt{\frac{12\pi e^2 N}{m^* v_F^2}}\ {\rm for}\ I,\\
    \sqrt{\frac{4\pi e^2 N}{T}}\ {\rm for}\ II,
    \end{array}
    \right.
\end{equation}
where $N$ is the density of carriers. Note that $\varepsilon_t$ and
$\varepsilon_l$ are of the first order in $e^2$ and can be used with
the Lifshitz formula. The functions $f_{t,l}(x)$ are different for
each class and can be found in the textbooks (see, for example,
\cite{LP10}). For the class I they are
\begin{eqnarray}
\nonumber  f_t(x) &=& \frac{3x}{2}\left[x-\frac{1}{2}(x^2-1)\ln\frac{x+1}{x-1}\right], \\
\label{fI}  f_l(x) &=&
3x^2\left[-1+\frac{x}{2}\ln\frac{x+1}{x-1}\right].
\end{eqnarray}
For the class II these functions are
\begin{equation}\label{fII}
    f_t(x)=-F(x),\ \ \ f_l(x)=-2x^2\left(1+F(x)\right),
\end{equation}
where $F(x)$ is defined as
\begin{equation}\label{Fdef}
    F(x)=\frac{x}{\sqrt{\pi}}\int\limits_{-\infty}^{\infty}dz\frac{e^{-z^2}}{z-x}.
\end{equation}

The dielectric functions (\ref{etl}) can be generalized to the case
of finite relaxation frequency $\gamma$. For the transverse
component $\varepsilon_t(\omega,k)$ this is an easy task since the
collision integral in the relaxation time approximation gives the
same result as (\ref{etl}) but with the finite value of the
relaxation frequency $\gamma$.

Different procedure has to be used to account for the finite
relaxation time in $\varepsilon_l(\omega,k)$. The longitudinal field
influences the charge distribution. In this case the relaxation of
the perturbed distribution toward "equilibrium" will be to the local
state of charge imbalance and not to the uniform distribution
\cite{War60}. Mermin \cite{Mer70} used this idea to generalize
$\varepsilon_l(k,\omega)$ calculated in RPA to the finite relaxation
time:
\begin{equation}\label{Mermin}
    \tilde{\varepsilon}_l(\omega,k)=\varepsilon_0(\omega)+\frac{(1+i\frac{\gamma}{\omega})\left[\varepsilon_l
    (\omega+i\gamma,k)-\varepsilon_0(\omega)\right]}{1+i\frac{\gamma}{\omega}\frac{\varepsilon_l
    (\omega+i\gamma,k)-\varepsilon_0(\omega)}{\varepsilon_l(0,k)-\varepsilon_0(0)}},
\end{equation}
where we denote the dielectric functions with the finite relaxation
frequency as $\tilde{\varepsilon}_{t,l}$.

The dielectric functions $\tilde{\varepsilon}_{l,t}(k,\omega)$ are
defined for an infinite medium. The spatial dispersion close to the
body surface needs special attention. Strictly speaking one can
define the nonlocal dielectric function only for infinite body, but
in special cases of specular and diffuse reflection of electrons on
the surface of the body it is also possible to do. We will consider
here the case of specular reflection. An electron reflected from the
interface with vacuum cannot be distinguished from that coming from
a fictitious medium on the vacuum side. In this way the specular
condition continues the medium with the interface to the infinite
medium.

For actual evaluation of the force in the case of materials with the
spatial dispersion it is more convenient to use the surface
impedances instead of the dielectric functions \cite{Esq04}. These
impedances are connected with the dielectric functions by the
relations
\begin{eqnarray}
% \nonumber to remove numbering (before each equation)s
\label{Zs}  Z_s(q,\omega) &=& \frac{i\omega}{\pi c}\int\limits_{-\infty}^{\infty}
\frac{dk_z}{\frac{\omega^2}{c^2}\tilde{\varepsilon}_t-k^2}, \\
\label{Zp}  Z_p(q,\omega) &=& \frac{i\omega}{\pi
c}\int\limits_{-\infty}^{\infty}
\frac{dk_z}{k^2}\left[\frac{q^2}{\frac{\omega^2}{c^2}\tilde{\varepsilon}_l}+
\frac{k_z^2}{\frac{\omega^2}{c^2}\tilde{\varepsilon}_t-k^2}\right],
\end{eqnarray}
where the wave vector is ${\bf k}=(q_x,q_y,k_z)$ and the $z$-axis is
perpendicular to the body surface.

The reflection coefficients are expressed via the impedances as
\begin{equation}\label{refl}
    r_s(q,\omega)=-\frac{\omega/ck_0-Z_s}{\omega/ck_0+Z_s},\ \ \
    r_p(q,\omega)=\frac{ck_0/\omega-Z_p}{ck_0/\omega+Z_p},
\end{equation}
where $k_0=\sqrt{\omega^2/c^2-q^2}$. Now we can use the reflection
coefficients (\ref{refl}) in the Lifshitz formula. This formula is
usually presented via the imaginary Matsubara frequencies
$\omega\rightarrow i\zeta_n=i2\pi Tn/\hbar$:
\begin{equation}\label{Lif}
    F(a,T)=\frac{T}{\pi}\sum\limits_{\mu=s,p}\sum\limits_{n=0}^{\infty}{}^{\prime}
    \int\limits_0^{\infty}dqq|k_0|\frac{ r_{\mu 1}r_{\mu
    2}e^{-2|k_0|a}}{1-r_{\mu 1}r_{\mu
    2}e^{-2|k_0|a}},
\end{equation}
where $r_{\mu i}=r_{\mu i}(q,\zeta_n)$ is the reflection coefficient
of the body $i$ ($i=1,2$) for the polarization $\mu$ ($\mu=s,p$).

Let us consider the influence of the spatial dispersion on the
force. When $T$ is around room temperature typical values of the
parameters are $\gamma\sim 10^{13}-10^{14}\;rad/s$, $k\sim 1/a$, and
$v\lesssim 10^{6}\;m/s$. Then the natural value of $x$ in
(\ref{xdef}) is large, $|x|\gg 1$, for most of the materials in the
interesting distance range 10-1000 nm. In this limit both dielectric
functions become local:
\begin{equation}\label{local}
    \tilde{\varepsilon}_t=\tilde{\varepsilon}_l(k,i\zeta_n)=
    \varepsilon_0(i\zeta_n)+\frac{\omega_p^2}{\zeta_n\left(\zeta_n+\gamma\right)},
    \ \ \ n>0
\end{equation}
This is true, however, for $n>0$. In the case $n=0$ or
$\zeta\rightarrow 0$ the result is different:
\begin{equation}\label{nzero}
    \tilde{\varepsilon}_t(k,i\zeta) =
    \varepsilon_0+\frac{\omega_p^2}{\zeta\gamma},\ \ \
    \tilde{\varepsilon}_l(k,i\zeta) =
  \varepsilon_0+\frac{k_D^2}{k^2},
\end{equation}
where $\varepsilon_0=\varepsilon_0(0)$. In the real frequency domain
this limit is realized for $\omega\ll \gamma$. As one can see
$\tilde{\varepsilon}_l$ demonstrates the nonlocal character.

We can conclude that at room temperature all the $n>0$ terms in
(\ref{Lif}) can be calculated with the local dielectric functions.
The $n=0$ term has to be corrected to take into account the
nonlocality. Calculating the impedances (\ref{Zs}) and (\ref{Zp})
with the functions (\ref{nzero}) and substituting them into the
reflection coefficients (\ref{refl}) one finds
\begin{equation}\label{rp0}
r_s\rightarrow 0,\ \ \
r_p=\frac{\varepsilon_{0}\sqrt{q^2+k_D^2/\varepsilon_0}-q}
{\varepsilon_{0}\sqrt{q^2+k_D^2/\varepsilon_0}+q}.
\end{equation}
Here $r_s$ is zero because in the static limit $s$-polarized field
is reduced to pure magnetic field, which penetrates the nonmagnetic
material. One can see that $r_p$ interpolates between a good metal
and a pure dielectric. Really, important values of $q$ are $q\sim
1/a$, then for $a\ll l_D=\sqrt{\varepsilon_0}/k_D$ the reflection
coefficient $r_p=(\varepsilon_0-1)/(\varepsilon_0+1)$ is as for a
dielectric with the permittivity $\varepsilon_0$. In the opposite
limit $r_p=1$ corresponds to good metals.

In the local theory the $n=0$ term in the Lifshitz formula
(\ref{Lif}) can be presented in the Lifshitz form \cite{DLP}
\begin{equation}\label{nzLif}
    F_0(a,T)=\frac{T}{16\pi a^3}\int\limits_0^{\infty}dx\frac{x^2 Re^{-x}}{1-R e^{-x}},
\end{equation}
where $R=r_{p1}r_{p2}$ is just a constant. For example, for two
dielectrics with permittivities $\varepsilon_1$ and $\varepsilon_2$
it is
\begin{equation}\label{R_loc}
    R=\frac{(\varepsilon_1-1)(\varepsilon_2-1)}{(\varepsilon_1+1)(\varepsilon_2+1)}.
\end{equation}

In the more general theory, which takes into account the spatial
dispersion of the materials, the $n=0$ term can be presented in the
same form (\ref{nzLif}) but now $R$ is not a constant but a function
of $q$ as follows from (\ref{rp0}). Introducing for each material
the parameter $\xi_i=2a/l_{iD}$ ($i=1,2$) we can present
$R=r_{p1}r_{p2}$ as
\begin{equation}\label{R_nonloc}
    R(x)=\frac{\varepsilon_1\sqrt{x^2+\xi^2_1}-x}{\varepsilon_1\sqrt{x^2+\xi^2_1}+x}\cdot
    \frac{\varepsilon_2\sqrt{x^2+\xi^2_2}-x}{\varepsilon_2\sqrt{x^2+\xi^2_2}+x}.
\end{equation}
From (\ref{nzLif}) and (\ref{R_nonloc}) we can reproduce the main
formula (34) in \cite{Pit08}. For that in the large distance limit
we can consider the second body as rarefied $\varepsilon_2-1\ll 1$,
$\xi_2\rightarrow 0$ and calculate the atom-body potential as
$V=-F_0(a,T)$.

Our analysis of the Casimir-Lifshitz force for poor conductors at
room temperature can be applied for the description of a recent
experiment \cite{Che07a,Che07b}, the result of which did not found a
reasonable explanation yet. In this experiment a p-type silicon
membrane with the carrier density $N\approx 5\cdot 10^{14}\ cm^{-3}$
was excited by the laser light. The density of photogenerated
carriers was varied in the range $N_{ph}=(1.4-2.1)\cdot 10^{19}\
cm^{-3}$. The force difference in the presence and in the absence of
laser light was measured with an atomic force microscope (AFM) as a
function of the distance between the  membrane and the sphere at the
end of the cantilever. It was found that the experimental results
agree well with the dielectric membrane rather than with the
semiconducting one. On this basis a controversial conclusion was
made that one has to disregard the finite conductivity of silicon
when the Casimir-Lifshitz force is evaluated.

The controversy manifests itself in the local variant of the theory
as discontinuity of the $n=0$ term in the Lifshitz formula. For a
dielectric (body 1) and metal (body 2) this term is given by
(\ref{nzLif}) with $R=(\varepsilon_1-1)/(\varepsilon_1+1)$. Even for
infinitely small conductivity of the dielectric this term has to be
calculated with $R=1$ and the result will coincide with the $n=0$
term for two metals. The nonlocality smooth out this discontinuity
allowing continuous transition from dielectric to metal in
accordance with the common sense.

\begin{figure}[ptb]
\begin{center}
\includegraphics[width=0.45\textwidth]{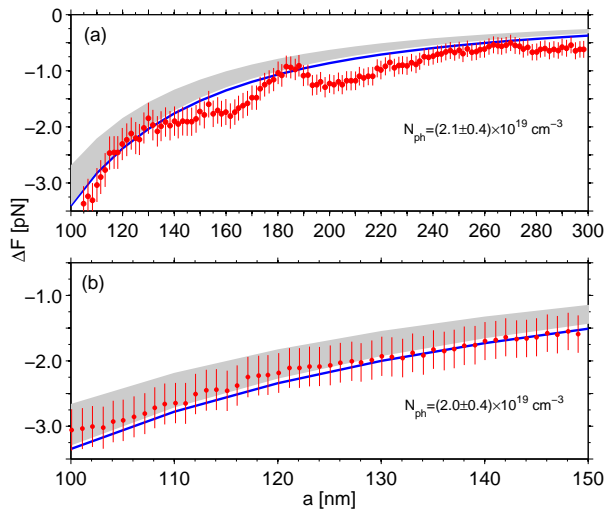}
\vspace{-0.3cm} \caption{(Color online) Difference of the forces on
the low-doped Si membrane in the presence and in the absence of
photogenerated carriers (dots with bars). The gray stripes are
predictions of this work accounted for errors in $N_{ph}$. The blue
curves are predictions based on pure dielectric Si for the central
values of $N_{ph}$. (a) and (b) correspond to different laser powers
exciting the carriers (see \cite{Che07b}).} \label{fig1}
\vspace{-0.9cm}
\end{center}
\end{figure}

Figure \ref{fig1} shows the experimental data for two densities of
the photogenerated carriers Note that $N_{ph}$ in both cases are the
same within the errors. Comparing figures \ref{fig1}(a) and
\ref{fig1}(b) one can see that the experiment can hardly distinct
between the nonlocal theory and the local theory with zero
conductivity of Si.

The controversies of the local theory are closely related with the
behavior of the entropy at $T\rightarrow 0$. Let us check the Nernst
heat theorem in the nonlocal case. At low temperatures we cannot use
the dielectric functions (\ref{local}) and (\ref{nzero}) since the
variable $|x|$ now is not large. Actually, when $T$ is sufficiently
small, the opposite limit is realized $|x|\ll 1$, for which the
nonlocal effects are strong. This is because $\gamma(T)$ decreases
with $T$ faster than linearly. Important imaginary frequencies
contributing to the temperature dependent part of the free energy
are $\zeta\sim 2\pi T/\hbar$. Now the nonlocality is important for
many terms in the sum (\ref{Lif}). In this limit the dielectric
functions are
\begin{equation}\label{etlT}
\tilde{\varepsilon}_t(i\zeta,k)=\varepsilon_0(i\zeta)+\frac{\alpha\omega_p^2}{\zeta
v k},\ \ \
\tilde{\varepsilon}_l(i\zeta,k)=\varepsilon_0(i\zeta)+\frac{k_D^2}{k^2}.
\end{equation}
Here we suppressed the index $n$ and $\alpha$ is $3\pi/4$ for the
class I and $\sqrt{\pi}$ for II. Note that $\gamma$ falls out from
the result; its role plays the Landau damping frequency $vk$.

%===================================================================================================
%===================================================================================================

Luckily there is no need to calculate the free energy with
(\ref{etlT}); all the work was done before. For the class I
materials the calculations were performed in \cite{Sve05} but the
distance dependent leading term was presented in \cite{Sve06}:
\begin{equation}\label{Free}
    {\cal F}\approx-0.0193\;T^2\frac{a\omega_p^2}{\hbar v_F c^2},
    \ \ \ T\ll\frac{\hbar v_F c^2}{12\pi^2\omega_p^2 a^3}.
\end{equation}
The spatial dispersion changes the behavior of the free energy with
$T$: instead of linear as in the local theory \cite{Bez02} it
becomes quadratic. It ensures the right behavior of the entropy
$S\sim +T$.

Different situation is realized for the class II materials. In this
case the density of carriers $N(T)$ is a function of temperature.
Due to presence of the energy gap for these materials at
$T\rightarrow 0$ the dependence is exponential $N(T)\sim
e^{-\Delta/T}$, where $\Delta $ is the gap. Because both parameters
$\omega_p^2$ and $k_D^2$ are proportional to $N(T)$ their effect in
the dielectric functions (\ref{etlT}) is exponentially small. This
conclusion is also true for the hopping mechanism of conductivity.
In this case the effective density of charges, which are able to
move, is $\sim e^{-\Delta/T}$ \cite{Lin79}. In the limit
$T\rightarrow 0$ the class II materials become pure dielectric. For
dielectrics the entropy disappears with $T$ as $S\sim T^2$
\cite{Gey05}.

In conclusion, the Lifshitz formula was generalized to materials,
which cannot be described as local media. The spatial dispersion
naturally resolves paradoxes appearing in the theory for local
materials with finite conductivity.

When this work was finished a preprint \cite{Dal08} appeared where
the Lifshitz theory was generalized to semiconductors. Ideologically
the approach is similar to ours but less general and technically
different.

Fruitful discussion with L. P. Pitaevskii is appreciated.

 %%%%%%%%%%%%%%%%%%%%%%%%%%%%%%%%%%%%%%%%%%%%%%%%%%%%%%%%%%%%%%%%%%%%%%%%%%%%%%%%%%%%%%%%%%%%%%%%


\begin{thebibliography}{99}

\bibitem{Cas48} H.B.G. Casimir, Proc. K. Ned. Akad. Wet. {\bf 51}, 793 (1948).

\bibitem{Lif55} E.M.~Lifshitz, Dokl. Akad. Nauk SSSR {\bf 100}, 879 (1955).

\bibitem{DLP}  I.E.~Dzyaloshinskii, E.M.~Lifshitz and L.P.~Pitaevskii,
Advances in Physics {\bf 38}, 165 (1961).

\bibitem{LP9} E. M. Lifshitz and L. P. Pitaevskii, {\it Statistical
Physics} (Pergamon Press, Oxford, 1980) Pt. 2.

\bibitem{Cap07} F. Capasso, J. N. Munday, D. Iannuzzi, and H. B.
Chan, IEEE J. Sel. Top. Quantum Electron. {\bf 13}, 400 (2007).

\bibitem{Bos00} M.~Bostr\"{o}m and B.E.~Sernelius, Phys. Rev. Lett.
\textbf{84}, 4757 (2000).

\bibitem{Bez02} V. B. Bezerra, G. L. Klimchitskaya, and V. M. Mostepanenko,
Phys. Rev. A {\bf65 }, 052113 (2002).

\bibitem{Nernst} G. L. Klimchitskaya and V. M. Mostepanenko, Phys.
Rev. E {\bf 77}, 023101 (2008); J. S. H{\o}ye, I. Brevik, S. A.
Ellingsen, and J. B. Aarseth, Phys. Rev. E {\bf 77}, 023102 (2008).

\bibitem{Gey05} B. Geyer, G. L. Klimchitskaya, and V. M. Mostepanenko,
Phys. Rev. D {\bf 72}, 085009 (2005).

\bibitem{Sve03} V. B. Svetovoy and M. V. Lokhanin, Phis. Rev. A {\bf
67}, 022113 (2003).

\bibitem{Ser05} Bo E. Sernelius, Phys. Rev. B {\bf 71}, 235114
(2005); Phys. Rev. B {\bf 75}, 036102 (2007).

\bibitem{Sve05} V.B.~Svetovoy and R.~Esquivel,
Phys. Rev. E \textbf{72}, 036113 (2005).

\bibitem{Pit08} L. P. Pitaevskii, arXiv: 0801.0656.

\bibitem{Bar75} Y. S. Barash and V. L. Ginzburg, Usp. Fiz. Nauk {\bf 116}, 5 (1975) [Sov. Phys. Uspekhi
{\bf 18}, 305 (1975)].

\bibitem{Kli07} G. L. Klimchitskaya and V. M. Mostepanenko, Phys.
Rev. B {\bf 75}, 036101 (2007).

\bibitem{Esq04} R. Esquivel and V. B. Svetovoy, Phys. Rev. A {\bf 69}, 062102
(2004).

\bibitem{Ash76} N. W. Ashcroft and N. D. Mermin, {\it Solid State
Physics} (Thomson Learning, Toronto, 1976).

\bibitem{Gor80} V. D. Gorobchenko and E. G. Maksimov, Usp. Fiz. Nauk {\bf 130}, 65 (1980)
[Sov. Phys. Usp. {\bf 23}, 35 (1980)].

\bibitem{LP10} E. M. Lifshitz and L. P. Pitaevskii, {\it Physical
Kinetics} (Pergamon Press, Oxford, 1981).

\bibitem{Sve06} V.B.~Svetovoy and R.~Esquivel, J. Phys. A: Math. Gen.
\textbf{39}, 6777 (2006).

\bibitem{War60} J. L. Warren and R. A. Ferrell, Phys. Rev. {\bf
117}, 1252 (1960).

\bibitem{Mer70} N. D. Mermin, Phys. Rev. B {\bf 1}, 2362 (1970).

\bibitem{Che07a} F. Chen, G. L. Klimchitskaya, V. M. Mostepanenko, and U. Mohideen,
Opt. Express {\bf 15}, 4823 (2007).

\bibitem{Che07b} F. Chen, G. L. Klimchitskaya, V. M. Mostepanenko, and U.
Mohideen, Phys. Rev. B {\bf 76}, 035338 (2007).

\bibitem{Lin79} M. E. Lines, Phys. Rev. B {\bf 19}, 1183 (1979).

\bibitem{Dal08} D. A. R. Dalvit and S. Lamoreaux, arXiv: 0805.1676.




\end{thebibliography}
\end{document}